\documentclass[%
 reprint,
 amsmath,amssymb,
prl,
]{revtex4-1}

\usepackage{graphicx}
\usepackage{dcolumn}
\usepackage{bm}
\usepackage{hyperref}
\usepackage[normalem]{ulem} 

\begin{document}


\title{First-principles Equation of State and Shock Compression Predictions of Warm Dense Hydrocarbons}

\author{Shuai Zhang}
\email{shuai.zhang01@berkeley.edu}
\author{Kevin P. Driver}
\affiliation{Department of Earth and Planetary Science, University of California, Berkeley, California 94720, USA (Current address: Lawrence Livermore National Laboratory, Livermore, California 94550, USA)}
\author{Fran\c{c}ois Soubiran}
\affiliation{Department of Earth and Planetary Science, University of California, Berkeley, California 94720, USA}
\author{Burkhard Militzer}
\email{militzer@berkeley.edu}
\affiliation{Department of Earth and Planetary Science, University of California, Berkeley, California 94720, USA}
\affiliation{Department of Astronomy, University of California, Berkeley, California 94720, USA}

\date{\today}

\begin{abstract}
  We use path integral Monte Carlo and density functional molecular
  dynamics to construct a coherent set of equation of state for a
  series of hydrocarbon materials with various C:H ratios (2:1, 1:1,
  2:3, 1:2, and 1:4) over the range of $0.07-22.4$ g~cm$^{-3}$ and
  $6.7\times10^3-1.29\times10^8$ K.  The shock Hugoniot curve derived
  for each material displays a single compression maximum
  corresponding to $K$-shell ionization.  For C:H=1:1, the compression
  maximum occurs at 4.7-fold of the initial density and we show
  radiation effects significantly increase the shock compression ratio
  above 2 Gbar, surpassing relativistic effects.  The single-peaked
  structure of the Hugoniot curves contrasts with previous work on
  higher-$Z$ plasmas, which exhibit a two-peak structure corresponding
  to both $K$- and $L$-shell ionization.  Analysis of the electronic
  density of states reveals that the change in Hugoniot structure is
  due to merging of the $L$-shell eigenstates in carbon, while they
  remain distinct for higher-$Z$ elements.
  Finally, we show that the
  isobaric-isothermal linear mixing rule for carbon and hydrogen EOSs
  is a reasonable approximation with errors better than
  1\% for stellar-core conditions.
\end{abstract}

\maketitle




\textit{Introduction.}
Hydrocarbon ablator materials are of primary importance for
laser-driven shock experiments, such as those central to the study of
inertial confinement fusion
(ICF)~\cite{Betti2016,Meezan2016,Goncharov2016} and the measurement of
high energy density states relevant to giant
planets~\cite{Guillot2005} and stellar
objects~\cite{Wallerstein1997}. Accurate knowledge of the equation of
state (EOS) of the hydrocarbon ablator is essential for optimizing
experimental designs to achieve desired density and temperature states
in a target. Consequently, a number of planar-driven shock wave
experiments have been performed on hydrocarbon materials, including
polystyrene
(CH)~\cite{Hauver1964,Dudoladov1969,Lamberson1972,Marsh1980,Nellis1984,Kodama1991,Bushman1996,Koenig1998,Cauble1997,Cauble1998,Koenig2003,Ozaki2005,Hu2008,Ozaki2009,Barrios2010,Shu2010,Shang2013,Shu2015},
glow-discharge polymer (GDP)~\cite{Barrios2012,Huser2013,Huser2015,Moore2016,Hamel2012},
and
foams~\cite{Bolkhovitinov1978,Koenig1999,Takamatsu2003,Koenig2005}, to
measure the EOS. The highest pressure achieved among these experiments
is 40 Mbar~\cite{Cauble1997,Cauble1998}, which is yet not high enough to probe
the effects of $K$-shell ionization on the shock Hugoniot
curve. 
Since the first X-ray scattering results on CH at above 0.1 Gbar 
(1 Gbar=100 TPa)~\cite{Kraus2016},
ongoing, spherically-converging shock experiments using the
Gbar platform at the National Ignition
Facility (NIF)~\cite{Swift2011,Doppner2014,Kritcher2014,Kritcher2016,Nilsen2016} and the OMEGA laser~\cite{Nora2015} will
extend measurements of the shock Hugoniot curve of polystyrene 
to pressures above 0.35 Gbar and into
the $K$-shell ionization regime~\cite{CHGbarExpts}.


These experiments provide an important benchmark for the theoretical
community working on models for EOSs of warm dense matter (WDM). EOS
tables, such as SESAME~\cite{sesame} and QEOS~\cite{More1988},
which are largely based on variations of Thomas-Fermi (TF) models, are
outdated and yet still often used in hydrodynamic simulations for the
design of shock experiments.  There have been ongoing efforts to
develop efficient first-principles methods for WDM that maintain an
accurate treatment of the many-body and shell-ionization effects that
TF method neglects~\cite{FCWDM}. Standard Kohn-Sham density functional
theory molecular dynamics (DFT-MD) is a suitable method for low and
intermediate temperatures, but becomes computationally intractable
beyond temperatures of 100 eV, where $K$-shell ionization becomes
important in mid-$Z$ elements.

The efficiency limitation of thermal DFT-MD has largely been addressed
by the development of orbital-free (OF)~\cite{Lambert2006} and
average-atom~\cite{Purgatorio2006} approximations. Indeed, a number of
calculations employing both DFT-MD and OF-DFT have been used to study
the EOS of hydrocarbon materials, including
polystyrene~\cite{Mattsson2010,Wang2011,Lambert2012,Hu2014,Hu2015,Hu2016}, polyethylene \cite{Knyazev2015}, 
and GDP~\cite{Huser2015,Colin2016}, in the WDM regime. The highest
density and temperature simulations among these calculations have been
performed with OF-DFT, up to 100 g~cm$^{-3}$ and 345
eV~\cite{Hu2015}. At low temperatures, these approximate DFT-based
simulations predict the shock Hugoniot curve in good agreement with
experiments. However, there are important limitations to their
accuracy. OF-DFT replaces the orbital-based kinetic energy functional
with a density-based TF functional and, therefore, is also unable to
account for shell ionization effects~\cite{PhysRevB.94.094109}. On the other hand, DFT-based
average atom methods only compute shell structure for an average ionic
state and, subsequently, it is not well-suited for studies of
compounds.

As an alternative to DFT-based methods, the path integral Monte Carlo (PIMC)
method~\cite{Ceperley1995,Ce96} offers an approach to explicitly treat
all the many-body and ionization effects as long as a suitable nodal structure is employed. Early
developmental work established the accuracy of the method for
fully-ionized
hydrogen~\cite{Pierleoni1994,PhysRevE.63.066404} and
helium~\cite{Militzer2009} plasmas using free-particle nodes. 
In recent works, we have further developed
free-particle~\cite{Driver2012} and
localized~\cite{Militzer2015Silicon} nodal structures, which has allowed us to compute first-principles
EOSs across a wide range of density-temperature regimes for heavier, first- and second-row,
elements.

In this work, we combine low-temperature DFT-MD data with
high-temperature PIMC data to compute coherent EOSs for several
hydrocarbon materials across a wide density-temperature range. We aim
to provide a highly accurate theoretical benchmark for the shock
Hugoniot curves, which can help guide hydrodynamic target
designs and interpret ongoing Gbar spherically-converging shock
experiments in the WDM regime, particulary where $K$-shell ionization
effects arise. While such state-of-the-art shock experiments maintain
exquisite control over many experimental parameters, difficulties can
remain in the interpretation of results due to insufficient knowledge of
the opacity in the density unfolding process of radiographic
measurements~\cite{Swift2011}, preheating effects, or shock uniformity
and stability.  Comparing both theoretical and experimental benchmarks
can offer great insight into narrowing down and eliminating future
sources of error.

\textit{Methods.} 
We consider five different C:H ratios of 2:1, 1:1, 2:3, 1:2, and 1:4,
in order to cover the full range of interest in future shock
experiments. Depending on the C:H ratio, our simulation cells
contain between 30 and 50 nuclei, as well as between 100 and 130 electrons.
In order to eliminate the finite-size effects at low temperatures, 
we use four times larger cells at temperatures up to 2$\times10^5$ K. 
Above 2$\times10^4$ K and 400 GPa, the Hugoniot curves derived with the 
small- and large-cell results are indistinguishable.

Using the {\footnotesize CUPID} code~\cite{militzerphd}, we perform
PIMC simulations within the fixed node approximation~\cite{Ceperley1991,Ce96}.
Similar to the PIMC simulations of
hydrogen~\cite{Pierleoni1994,PhysRevLett.76.1240,PhysRevE.63.066404,PhysRevLett.87.275502,PhysRevLett.104.235003,PhysRevLett.85.1890,PhysRevB.84.224109,CTPP:CTPP2150390137,Militzer20062136},
helium~\cite{Militzer2009,PhysRevLett.97.175501}, H-He mixtures \cite{Militzer2005}, carbon~\cite{Driver2012,Benedict2014C},
nitrogen~\cite{Driver2016Nitrogen}, oxygen~\cite{Driver2015Oxygen},
neon~\cite{Driver2015Neon}, and water~\cite{Driver2012}, we employ
a free-particle nodal structure. 
We enforce fermion nodes at a small imaginary time 
interval of 1/8192 Hartree$^{-1}$ (Ha$^{-1}$) while pair density
matrices~\cite{pollock1988,pdm} are evaluated in larger 
step of 1/1024 Ha$^{-1}$~\cite{PhysRevLett.85.1890}.

DFT-MD simulations use the Vienna \textit{Ab initio} Simulation
Package ({\footnotesize VASP})~\cite{kresse96b} and
exchange-correlation functionals within the local density
approximation (LDA)~\cite{Perdew81,Ceperley1980}.  We use all-electron
projector augmented wave (PAW) pseudopotentials~\cite{Blochl1994} with
a 1.1 and a 0.8 Bohr radius core for carbon and hydrogen,
respectively.  We use a plane wave basis with 2000 eV cutoff, the
$\Gamma$-point for sampling the Brillouin zone, a MD timestep of
0.05-0.2 fs, and a $NVT$ ensemble controlled with a Nos\'e
thermostat~\cite{Nose1984}.  Typical MD trajectories consist of more
than 1000 steps.  Longer simulations of up to 2 ps show that the
energies and pressures are converged. In order to put the DFT-MD
pseudopotential energies on the same all-electron scale as PIMC
calculations, we shifted all of our {\footnotesize VASP} energies by
$-$37.4243 Ha/C and $-$0.445893 Ha/H. These shifts were determined by
performing all-electron calculations for isolated C and H atoms with
the {\footnotesize OPIUM} code~\cite{opium}.

\begin{figure}[!htbp]
\centering\includegraphics[width=0.5\textwidth]{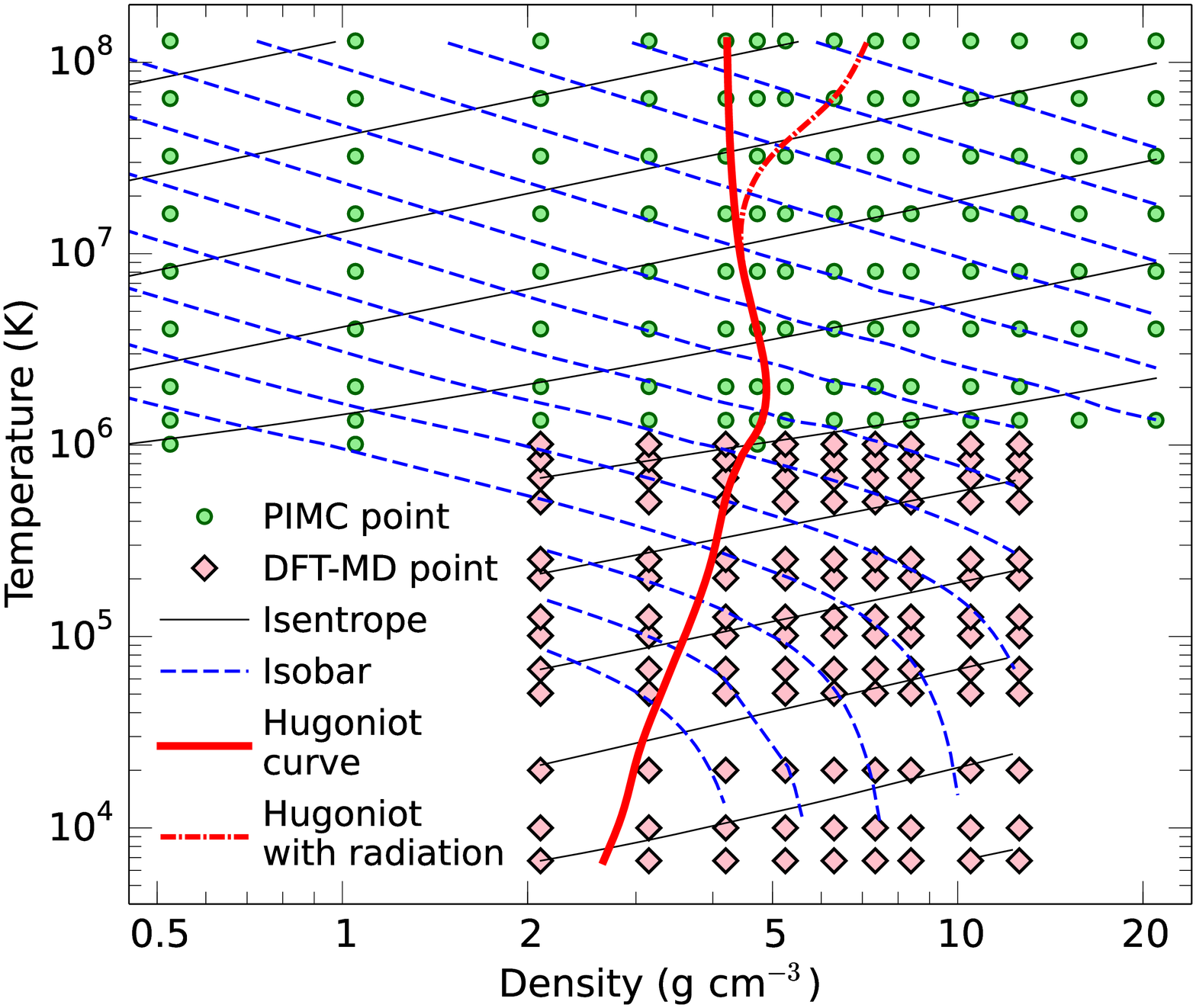}
\caption{\label{fig1} Temperature-density profile of the principal 
Hugoniot curve of polystyrene CH 
(initial density $\rho_0$=1.05 g$\,$cm$^{-3}$)
with and without radiation correction, 
obtained from the equation of state calculated in this work. 
Isobar and isentrope profiles are co-plotted. The symbols 
mark the simulation conditions.}
\end{figure}

\textit{Results and discussion.}
We performed DFT-MD at 6.7$\times10^3$-10$^6$ K and 2-12 times 
the ambient density of $\rho_\text{ambient}$ = 1.05 g$\,$cm$^{-3}$.
PIMC simulations were performed for 10$^6$-1.29$\times10^8$ K 
and a much wider density
range of 0.1-20 $\rho_\text{ambient}$
 since this method does not rely on
plane waves expansions nor pseudopotentials. 
At 10$^6$ K, the internal energy 
obtained from the two methods agreed to within 0.8 Ha/CH
while pressures agreed to within 2\%. We thus obtain a
coherent first-principles EOS table~\cite{supp}
 for hydrocarbon compounds over a wide
range of temperatures and densities 
and show that of polystyrene (CH) in Fig.~\ref{fig1}.

We then use the EOS to determine the Hugoniot curves, by computing the
$P$-$V$-$T$ conditions that satisfy the Hugoniot equation $(E-E_0) +
(P+P_0)(V-V_0)/2 = 0$, where $(E_0, P_0, V_0)$ and $(E, P, V)$ denote
the initial and final internal energies, pressures, and volumes in a
shock experiment, respectively.  The initial conditions were
determined based on thermo-physical and thermo-chemical data at 
1~bar~\cite{supp}. For instance, the density of polystyrene at
ambient is $\rho_0$=1.05~g/cm$^3$ \cite{NIST} and using the enthalpy
of combustion \cite{walters_2000} we determined $E_0=-38.3224$~Ha/CH.
The principal Hugoniot curve of polystyrene is plotted in
Fig.~\ref{fig1}.  Besides the dependence of shock velocity, the shock
compression is controlled by the excitation of internal degrees of
freedom, which increases the compression, and interaction effects,
which decrease it~\cite{PhysRevLett.97.175501}. With increasing
temperature and pressure, polystyrene is increasingly compressed until 
a maximum density of 4.9 g$\,$cm$^{-3}$ reached at 2.0$\times10^6$ K,
which corresponds to the excitation of $K$ shell electrons of carbon
ions, as we will explain below.

At temperatures above 10$^6$ K, radiation effects can no longer be
neglected. We re-construct the Hugoniot curve by considering the
contribution of an ideal black body radiation to the EOS via
$P_\textrm{photon}=4\sigma T^4/{3c}$ and
$E_\textrm{photon}=3P_\textrm{photon}V$, where $\sigma$ is the
Stefan-Boltzmann constant and $c$ is the speed of light in vacuum.
This is only an upper limit as the system is more likely to be a gray body with unknown efficiency.
With the radiation contribution, 
the Hugoniot curve shifts 
to significantly higher densities at above $10^7$ K and 
2 Gbar, while the $K$-shell compression peak remains
unchanged (see Fig.~\ref{fig1}).
This shift
can primarily be attributed to the photon contribution to the internal
energy.

\begin{figure}[!htbp]
\centering\includegraphics[width=0.5\textwidth]{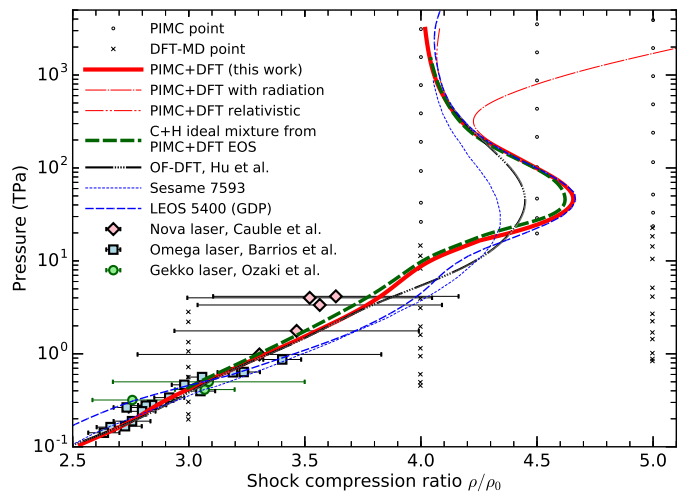}
\caption{\label{fig2}
  Pressure-compression profile of the principal Hugoniot curve of 
  polystyrene derived from this work in comparison 
with different theoretical and experimental methods:
 OF-DFT~\cite{Hu2015}, SESAME 7593~\cite{sesame},
  Nova~\cite{Cauble1997,Cauble1998}, Omega~\cite{Barrios2010,Hu2015} and
  Gekko~\cite{Ozaki2009} for CH, and  LEOS 5400~\cite{Sterne2016} for GDP.}
\end{figure}

A key result of shock experiments is the relation between the
compression ratio ($\rho/\rho_0$) and pressure. In Fig.~\ref{fig2}, we
compare our CH calculations with several experimental results and
theoretical predictions. Our results are in very good agreement with
experiments up to the highest pressure (4 TPa). We predict a maximum
compression ratio of 4.7 at 47 TPa, which is higher than results from
OF-DFT simulations ($\sim$4.4)~\cite{Hu2015} imply and SESAME 7593
($\sim$4.3)~\cite{sesame} but similar to predictions from 
a semi-analytical EOS model LEOS 5400 for GDP~\cite{Young1995,More1988,Sterne2016}. 
Still, all methods predict the
compression maximum to occur at very similar pressures. Our DFT-MD
results imply there is a small shoulder in the Hugoniot curve at
4.1-fold compression, $10^4$ GPa, and 6$\times 10^5$ K, which
separates in temperature the excitation regimes of the $K$ and $L$
shell electrons. Such a shoulder is absent from OF-DFT predictions
because this method predicts the ionization to occur gradually and
underestimates shell effects~\cite{PhysRevB.94.094109}. The shape of
SESAME Hugoniot curve is similar to that of OF-DFT, since it is largely
based on TF models.  The structure of the LEOS Hugoniot
curve is similar to the first-principles curve, but 
shows significant differences at low pressures due to the existence of oxygen in GDP.
Figure~\ref{fig2} also shows that the radiation
contributions dominate over relativistic effects even though both lead
to a compression ratio of 7 in the high-temperature limit, while the
limit for a non-relativistic gas is 4.

\begin{figure}[!htbp]
\centering\includegraphics[width=0.5\textwidth]{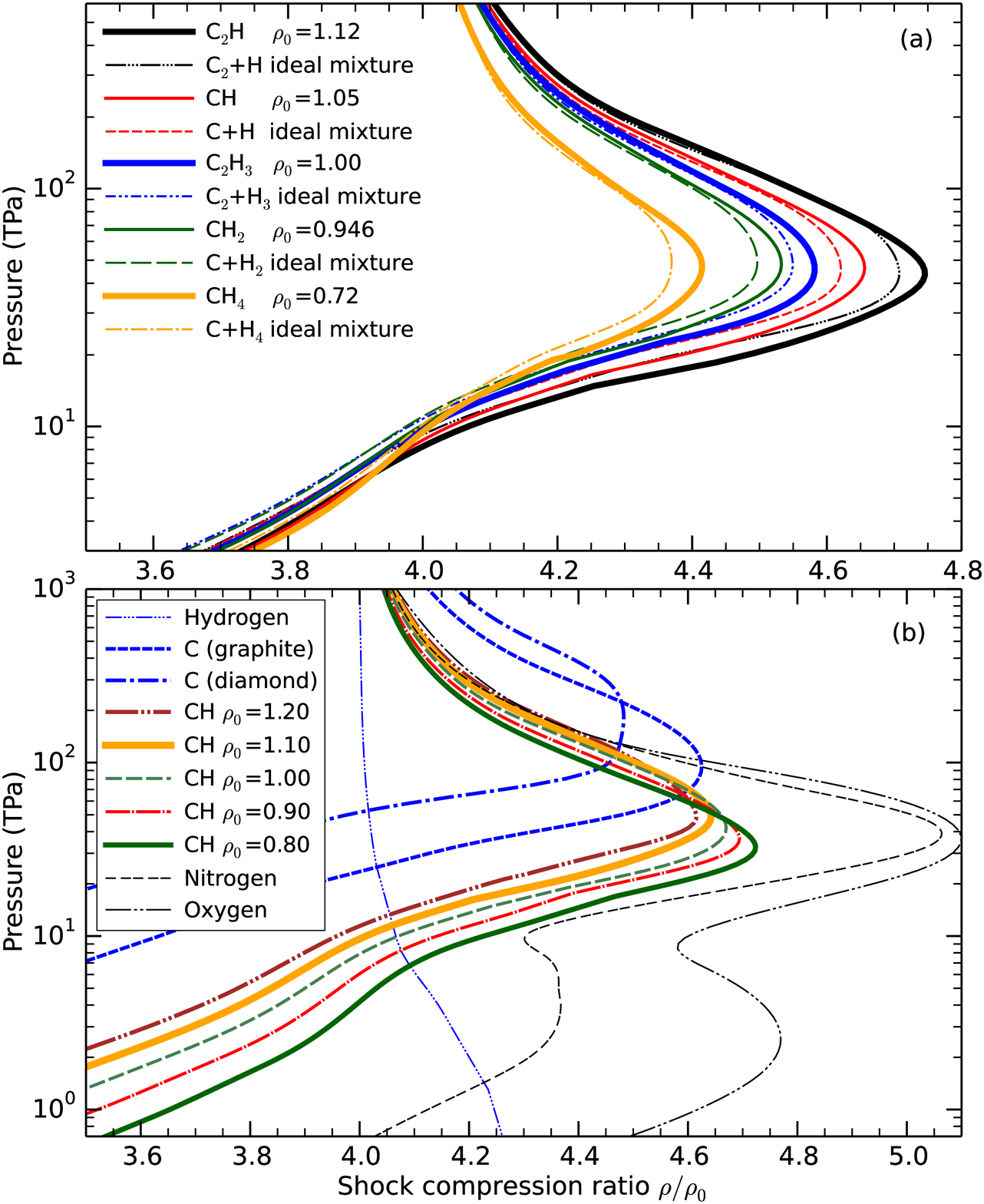}
\caption{\label{fig3} (a) Hugoniot curves of C-H compounds calculated
  from first principles in comparison with those from ideal mixtures
  of carbon and hydrogen.  (b) Hugoniot curves of polystyrene for
  different initial densities (in g$\,$cm$^{-3}$) and other materials.
}
\end{figure}

In Fig.~\ref{fig3}, we compare our predictions for the shock Hugoniot
curves of C$_2$H, CH, C$_2$H$_3$, CH$_2$, and CH$_4$ compounds. 
The initial conditions are determined by referring to representative 
hydrocarbon materials at ambient or cryogenic conditions~\cite{supp}.
For all C-H materials,
we find the compression maximum to occur at very similar pressures. At
the same time, we see a trend that lets the maximal compression ratio
gradually decrease from 4.7 to 4.4 as the hydrogen contents is
increased from C$_2$H to CH$_4$.
Our Hugoniot curves of graphite and
diamond do not follow this trend because the initial density of both
materials is much higher.
This implies that the particles interact
more strongly under shock conditions, which reduces the shock
compression ratio and shifts compression maximum towards higher
pressures~\cite{PhysRevLett.97.175501}. We also see this trend when we
compare the CH Hugoniot curves for different initial densities and
that of graphite and diamond with each other in Fig.~\ref{fig3}. 
The compression maximum appear at a
similar temperature ($\sim2\times10^6$ K) for all C-H compounds, graphite,
and diamond (see~\cite{supp}). This corresponds to the thermal ionization of 
the $K$ shell of carbon, as we will discuss later. The
magnitude of the shift in the compression maxima, $\sim$0.1, of the CH materials is
small compared to the deviations between predictions from various EOS
models in Fig.~\ref{fig2}.

When the EOS of mixtures needs to be derived for astrophysical
applications or to design shock wave experiments, one typically invokes
the ideal mixing approximation because the EOS of the fully interacting
systems, which we have computed here, is often not available. One
simply approximates the properties of the mixture as a linear
combination of the endmember properties at the same pressure and
temperature. For C-H mixtures, all interactions between C and H
particles are thus neglected. Furthermore, the ionization fraction of carbon
atoms in the C-H mixtures is set equal to the ionization fraction of
carbon at the same conditions. The presence of hydrogen does not
affect the ionization of carbon atoms and vice versa. 

For a mixture of heavier species and metallic hydrogen at conditions
in gas giant interiors, it has been shown that a linear mixing
approximation was not accurate~\cite{Soubiran2016} while it worked
very well for molecular H$_2$-H$_2$O mixtures in ice giant
envelopes~\cite{Soubiran2015}.  To test the validity of the linear
mixing approximation at higher temperatures, relevant for stellar
cores, we performed additional PIMC and DFT-MD simulations for pure H
and C and combined them with EOS tables from
Refs.~\cite{PhysRevB.84.224109,Benedict2014C}. Our results in
Figs.~\ref{fig2} and \ref{fig3} show that the linear mixing
approximation works exceptionally well for all C-H compounds. We only
see a small underestimation of the compression-ratio maximum of less
than 1\% ($\sim$0.035). Under these conditions, the shock compression
is controlled by the ionization equilibrium of $K$-shell electrons of
the C ions. This appears to be rather insensitive to whether a C ion
is surrounded by a C-H mixture or just by other C ions. We can thus
anticipate that the maximum uncertainty induced by using an ideal
mixing rule is below the 1\% level for stellar core conditions.

Figure~\ref{fig3} also shows that carbon and all C-H compounds exhibit only a
single compression maximum while nitrogen, oxygen, and
neon~\cite{Driver2015Neon} display two that have been attributed to
the excitation of $K$- and $L$-shell electrons. The fact that carbon
materials do not show the lower $L$-shell compression maximum of $\sim$3
TPa requires further investigation. 

\begin{figure}[!htbp]
\centering\includegraphics[width=0.5\textwidth]{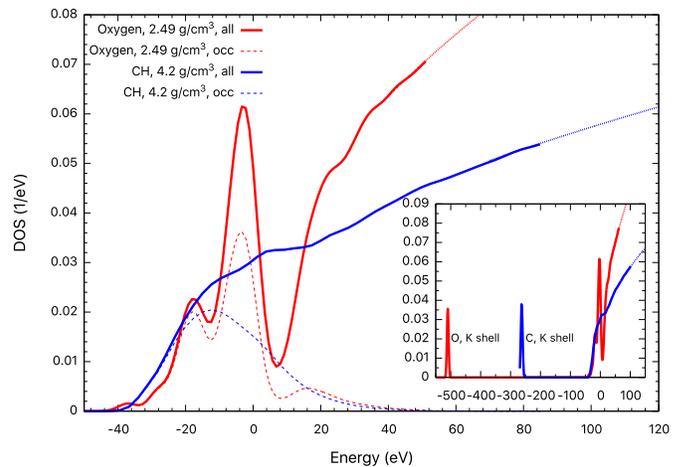}
\caption{\label{fig4}
  Density of state of polystyrene in comparison with that of 
  oxygen at $10^5$ K. The dashed curves denote the occupied 
  states, and the dotted lines are extrapolations of the
  free-electron density of state. The Fermi energies are aligned
  at E=0 eV.
  }
\end{figure}

In order to better understand the difference in the Hugoniot curve
shapes, we compare the electronic density of states (DOS) in
Fig.~\ref{fig4} that we have derived from DFT-MD simulations of oxygen
and polystyrene at 4-fold compression and 10$^5$ K.  Both DOSs show an
isolated peak at low energy, which corresponds to the electrons in $K$
shells of oxygen and carbon. Their thermal ionization leads to a
pronounced compression maximum along the Hugoniot curve.  However,
while oxygen DOS shows another set of sharp peaks corresponding to the
$L$-shell, the eigenstates of polystyrene are even distributed and
partially merged with the continuum.  It is the excitation of
electrons in these well-defined $L$-shell states that leads to the
second compression maximum for oxygen, nitrogen, and neon.  For carbon
and hydrocarbons, the $L$-shell ionization is much more gradual and
already starts at much lower temperatures~\cite{Potekhin2005} 
than for oxygen. This does not lead to a well-defined compression peak but only
to the shoulder in Hugoniot curve that we have discussed in
Fig.~\ref{fig2}.

\textit{Conclusions.}  We performed the first entirely
first-principles determination of hydrocarbon mixtures in the WDM
regime by including all non-ideal effects. Based on PIMC and DFT-MD,
we obtained coherent sets of EOS over wide range of density and
temperature conditions and derived the shock Hugoniot curves of a
series of hydrocarbon materials. For polystyrene, we predict a maximum
shock compression ratio of 4.7 while earlier estimates range from 4.3
to 4.7. Our calculated Hugoniot curve agrees very well with
experimental measurements and provides guidance for the interpretation
of experiments on the Gbar platform at NIF. We observe a single
compression maximum for hydrocarbon materials while there are two
compression maxima in the Hugoniot curve of nitrogen, oxygen and neon.
We have shown that this difference is related to the properties of the $L$-shell
ionization, which is much more gradual for carbon. We found
that the linear isobaric-isothermal mixing approximation works very
well, resulting in a discrepancy in the density of CH of 1\% or less
under stellar core conditions.
This implies that it is sufficient to derive only accurate EOS tables
for the endmembers in order to provide a thermodynamic description of
deep stellar interiors.

\textit{Acknowledgments.}
  This research is supported by DOE grants DE-SC0010517 and DE-SC0016248.
FS and BM acknowledge partial support from DOE Grant DE-NA0001859. 
SZ is partially supported by the PLS-Postdoctoral Grant of the Lawrence Livermore National Laboratory.
  Computational support was provided by the Blue Waters sustained-petascale
  computing project (NSF ACI-1640776).
  We thank L. Benedict for helpful discussions.
  


%
%
\end{document}